\documentclass{jca}

\usepackage{amssymb}
\usepackage{amsthm}
\usepackage{times}

\newtheorem{definition}{Definition}

\begin{document}

\title{Cellular Automata as a Model of Physical Systems}
\author{Donny Cheung\inst{1} \and Carlos A. P{\'e}rez-Delgado\inst{2,3}}
\institute{Department of Computer Science, University of Calgary, Calgary, AB, T2N 1N4, Canada
%%\and Institute for Quantum Computing, University of Waterloo, Waterloo, ON, N2L 3G1, Canada
\and School of Computer Science, University of Waterloo, Waterloo, ON, N2L 3G1, Canada 
\and Department of Physics and Astronomy, University of Sheffield, Sheffield, S3 7RH, UK}

\maketitle

\begin{abstract}
Cellular Automata (CA), as they are presented in the literature, are abstract mathematical models of computation.
In this paper we present an alternate approach: using the CA as a model or theory of physical systems and devices. 
While this approach abstracts away all details of the underlying physical system, it remains faithful to the fact that there \emph{is} an underlying physical reality which it describes.
This imposes certain restrictions on the types of computations a CA can \emph{physically} carry out, and the resources it needs to do so.
In this paper we explore these and other consequences of our reformalization.
\end{abstract}

%=====================================================================
\section{Introduction}
%=====================================================================

In recent years there has been an ever increasing interest in quantum models of computation.
Quantum versions of traditional computing models such as the Turing machine \cite{deutsch85,bernstein1}, finite automata\cite{watrousqfa}, families of acyclic circuits \cite{yao}, as well as cellular automata \cite{sw04,watrous} have all been proposed.
All these \emph{quantizations} are of great importance. At the same time, there is a drawback that these quantizations are prone to suffer.

The problem is, that none of these are physical theories.
Concretely, a proper quantization takes a classical Newtonian theory of the natural world, such as Hamiltonian mechanics, and transforms it by adding \emph{quanta}, non-commuting observables, etc.
However, none of the models mentioned above are physical theories. Quantizing these, may, or may not, lead to a sensible model of how nature works.
This would depend on how physical the computation model was to begin with.

In this paper our concern is cellular automata. 
Despite CA being used to model natural phenomena, the model is not a theory of nature. 
As a brief example, consider how the update rule of a CA is stated to be applied to all cells instantaneously.
For this to happen, there would have to be a method for synchronizing cells across a potentially infinite lattice. 

Setting aside superluminal communication for a moment, it is clear that the CA needs a method for synchronization across lattice cells.
There are several \emph{algorithms} for CA synchronization already in the literature \cite{async}.
In this paper we are not concerned with the algorithms, or \emph{how} the CA cells synchronize.
Rather, we are concerned with the fact that they \emph{have to} synchronize, somehow, and that this synchronization requires physical resources.

As with any model of computation (as described above), a drawback of using the traditional CA definition as a starting point for a quantized version, is that at the end, one is left with a model that may or may not sensibly represent real physical systems. 
With all previously proposed models of QCA, it is indeed an important question when, and when not, they have analogous physical systems.

In this paper we  attempt to rephrase CA not as a mathematical construct but rather as a physical theory, or model, that describes and abstracts a family of real-world systems.
This will allow us to properly formalize well-known facts about implementations of CA.
For instance, the use of extra memory, that is the second lattice that is generally used to implement CA algorithms in traditional computers, is shown not to be an ``implementation detail'', but rather a necessary requirement of any physical CA device.

Ultimately, this reformalization of CA will constitute the classical foundation upon which we will construct our quantum CA model.
A paper with the details of this QCA formalization has been prepared in tandem with this paper, and has now appeared as \cite{cp07}.
In that paper we concentrate fully on the quantum model. We show that QCA in this formalization \emph{always} represent real quantum systems.

Here we concentrate on the reformalization of the classical CA, and some of the lessons we can learn from it.

%=====================================================================
\section{Closed Cellular Automata}\label{sec:CCA}
%=====================================================================

We recall the standard definition of the cellular automaton:
\begin{definition}
A cellular automaton (CA) is a 4-tuple $(L,\Sigma,\mathcal{N},f)$ consisting of a $d$-dimensional lattice of cells $L=\mathbb{Z}^d$ indexed by integers, a finite set $\Sigma$ of cell states, a neighbourhood scheme $\mathcal{N}$, which is a finite list of lattice vectors from $\mathbb{Z}^d$, and a local transition function $f:\Sigma^{\mathcal{N}}\rightarrow\Sigma$.
\end{definition}
A {\em configuration} is simply an assignment of states from $\Sigma$ to each cell in $L$.
Given a configuration, the transition function $f$ determines a new configuration after a single discrete time step.
For each cell $x$, the new state is given by applying $f$ to the current states of the cells in the list $x+\mathcal{N}$.

When we consider this definition of a CA as a mathematical object, the evolution of a CA with a given local transition function $f$ from an initial configuration is well-defined.
However, if we wish to use this to model physical systems, questions about synchronization arise: if we apply the transition function $f$ to a neighbourhood and update the state of a particular cell, we alter the input data for the neighbouring cells.
In order to address this, some external memory resource needs to be used to store these new states before the cells themselves can be updated.
For example, when implementing CA algorithms on traditional comuputers, a standard way to address this is to have a second lattice of memory cells which can store the new configuration as it is computed.
After the entire configuration is computed, we may update the state of each cell of the original lattice independently.\cite{weimar}
We can physically implement this strategy by having, for example, a lattice of cells with two registers, using cell states $\Sigma\times\Sigma$.
One register would be used to store the configuration, while the other register would be used as temporary storage.

There are two main points here that we would like to highlight.
First, in an actual physical device, there may not be a need for distinguishing two different types of memory.
While the two-register model we described above gives us a way to see how a physical cellular automata device can be used to intrinsically simulate the evolution of any CA, the device itself is capable of implementing a more general set of transitions.
Second, if we are to avoid synchronization issues in these physical devices without external interactions with outside resources, it is necessary to adopt an interaction-update model.
That is, at each time-step of the evolution of such a device, there has to be an interaction phase where the cells interact with each other according to some local rule, followed by an update phase, where each internal state of each cell is independently updated.
This requirement is independent of how the cells attempt to co-ordinate and synchronize their update procedures with one another: any such method requires this extra memory space, and tiered step approach.

We now consider a formal model which provides an abstraction for any evolution of a physical cellular automata device.
First, the \emph{interaction} phase will involve performing a local interaction function $f:\Sigma^{\mathcal{N}}\rightarrow\Sigma^{\mathcal{N}}$ on the neighbourhood of each cell.
This is a generalization of the act of computing the local transition function and storing the result in auxiliary memory.
Note that in this model, the local interaction function is a map from a neighbourhood to itself, rather than from a neighbourhood to a single cell.
In order to avoid the issues of synchronization, we must ensure that this function can be applied to a particular neighbourhood in a manner which does not interfere with its application to any other neighbourhood.
In our previous illustration, this was done by logically partitioning each cell into two separate memory registers, one of which is never overwritten.
For a more general picture, we introduce the notion of \emph{translation commutativity}.
\begin{definition}
Given a local interaction function $f:\Sigma^{\mathcal{N}}\rightarrow\Sigma^{\mathcal{N}}$ for a lattice $L$ and a neighbourhood scheme $\mathcal{N}$, we say that $f$ is translation commutative if for any two cells $x,y\in L$, $f_x$ and $f_y$ commute, where $f_x$ denotes the local interaction function $f$ applied to the neighbourhood $\mathcal{N}+x$.
\end{definition}
As long as the local interaction function satisfies this property, it may be applied to each cell of the lattice in any order, or even in parallel.
The evolution remains deterministic.
Finally, the \emph{update} phase will simply apply a local update function $g:\Sigma\rightarrow\Sigma$ to each cell.
This occurs after all of the local interaction functions have been applied.

We can encapsulate this into a formal model, which we have named \emph{Closed Cellular Automata} to emphasize the fact that we are considering cellular automata as closed physical systems, which do not interact with external resources of any type.
\begin{definition}
A Closed Cellular Automaton (CCA) is a 5-tuple $(L,\Sigma,\mathcal{N}\!,f,g)$ consisting of a lattice of cells $L=\mathbb{Z}^d$, a finite state space $\Sigma$, a neighbourhood scheme $\mathcal{N}$, a translation commutative local interaction function $f:\Sigma^{\mathcal{N}}\rightarrow\Sigma^{\mathcal{N}}$, and a local update function $g:\Sigma\rightarrow\Sigma$.
\end{definition}
The global transition function for each time step consists of two phases.
First, the interaction phase consists of applying $f$ to each cell of the lattice $L$.
Since $f$ is translation commutative, the transition functions may be applied in any order.
We may thus define the global interaction function as $F=\prod_{x\in L}f_x$.
For some physical devices,  these transition functions may happen simultaneously and in parallel.
However, this is not assumed by the definition of CCA, and it is not necessary for having a well-defined global interaction function $F$.
The interaction phase is followed by the update phase, which consists of applying $g$ to each cell.
We may also define the global update function $G=\prod_{x\in L}g_x$.
In each time step, the global transition function $GF$ is applied to the entire lattice.

In our discussion above, we have seen that any CA can be intrinsically simulated by a Closed CA.
However, we can also view the set of Closed CA as a subset of traditional CA.
To see this, we define the \emph{extended neighbourhood} of a cell to be the union of all neighbourhoods containing the cell.
In one time step, the new state of this cell can be completely determined by the current state of this extended neighbourhood by applying the local interaction function to each neighbourhood, and then applying the single-cell local update function.
This yields a CA transition function which maps the state of the extended neighbourhood to a single cell and gives the same global transition function as the original Closed CA.
The neighbourhood scheme for this CA is simply the extended neighbourhood of each cell of the Closed CA.
Thus, we can think of these physical cellular automata devices as being both physical systems which implement CA, and also as CA in their own right.

%=====================================================================
\section{Reversibility}\label{sec:Rev}
%=====================================================================

Reversibility is an important consideration in a model of closed physical cellular automata devices.
Since closed physical systems in general tend to undergo reversible evolution, reversible CA are useful for modelling such systems.
We consider a cellular automata to be reversible if the global transition function is a bijection on the set of configurations.
There are two ways to introduce reversibility into the Closed CA model.
First, we may simply consider Closed CA which intrinisically simulate a reversible CA.
Second, we may consider Closed CA whose own evolution is reversible, having global transition functions $GF$ which are bijections on the set of configurations.
If we wish to explore the question of whether Closed CA can be used to model closed physical systems with homogeneous evolution, where information or energy is not implicitly leaked into the environment, then we should consider the second option.
In particular, when we consider cellular automata models of quantum systems in Section \ref{sec:QCA}, we start with a reversible model of CCA.

The Closed CA model gives a natural way to introduce reversibility.
We define a reversible CCA as a CCA in which both the local interaction function $f$ and the local update function $g$ are reversible.
In order to reverse this evolution, we could apply the reverse functions in the opposite order, that is, to apply $g^{-1}$ to each cell, and then $f^{-1}$ to each neighbourhood.
This implements the global transition function $F^{-1}G^{-1}$, which is the reverse of the original global transition function $GF$.
However, we can also express the reverse of a reversible CCA as another reversible CCA.
In order to do this, we implement $G^{-1}(GF^{-1}G^{-1})=F^{-1}G^{-1}$, using $g^{-1}$ as the local update function, and $Gf^{-1}G^{-1}$ as the local interaction function.
Note that $Gf^{-1}G^{-1}$ acts trivially on any cell not in the neighbourhood which $f$ acts upon.
This allows us to express $Gf^{-1}G^{-1}$ as a reversible function acting only on a local neighbourhood.
Note that since any non-reversible local interaction function $f$ or local update function $g$ will implicitly erase information, they cannot yield reversible CCA.
Thus, this description encapsulates all CCA with reversible evolution.

We know that any CA can be intrinsically simulated by a CCA, as described in Section \ref{sec:CCA}.
Thus, we can also simulate any reversible CA using a CCA using this technique.
However, we can also simulate any reversible CA using a reversible CCA by adapting a technique described in \cite{cabook}.
Given a reversible CA with an alphabet $\Sigma$ over a lattice $L$, let $\sigma(x,t)$ denote the state of the cell $x\in L$ at time step $t\in\mathbb{Z}$, and let $C(t)$ denote the configuration of the entire lattice.
Let us suppose that we have local transition rules for both the forward and reverse versions of the CA, so that $\sigma(x,t)$ can be computed from the configuration of $x+\mathcal{N}$ at either time step $t-1$ or $t+1$.
We can simulate such a reversible CA using a reversible CCA over the lattice $L$ using the alphabet $\Sigma\times\Sigma$.

First, we need an addition relation on the original alphabet, $+:\Sigma\rightarrow\Sigma$, which can be given by using any bijection from $\Sigma$ to $\mathbb{Z}_{|\Sigma|}$.
Now, we may view a configuration of the reversible CCA as two configurations of the original reversible CA over the alphabet $\Sigma$.
At time $t$, the first configuration contains the current configuration $C(t)$, while the second configuration contains previous configuration $C(t-1)$.
For a given cell $x$, the local interaction function $f_x$ reads the current configuration of its neighbourhood from the first configuration, and adds $\sigma(x,t+1)-\sigma(x,t-1)$ to the state of the second configuration.
Note that because each $f_x$ reads input from the first configuration and adds to the second, it is a translation commutative function.
Since the second configuration initially contains $C(t-1)$, it will contain $C(t+1)$ after the interaction phase.
The local update function simply swaps the contents of the two configurations, so that the first configuration now contains $C(t+1)$ and the second contains $C(t)$.
Note that this construction requires both the forward and reverse local transition rules of the original reversible CA.

Since we are proposing that reversible CCA be thought of as a model for closed physical implementations of cellular automata devices, we should also consider the question of whether reversible CA can be also generally thought of as closed physical systems.
However, the shift-right CA over a one-dimensional lattice, in which $\sigma(x,t+1)=\sigma(x-1,t)$, cannot be expressed in terms of a reversible CCA or, in fact, any system of local operations.
To see this, consider any circuit, over an infinite one-dimensional lattice, which does effect the shift-right operation using only local operations---ones which act on a finite neighbourhood.
We may suppose that at any time step, an infinite number of gates may be operating throughout the lattice in parallel, but we may insist that the \emph{depth} of the circuit be finite, so that the shift-right global operation $F$ can be decomposed as $F=f_{n}f_{n-1}\ldots f_{2}f_{1}$, where each $f_{k}$ is the product of potentially infinitely many local gates on disjoint local neighbourhoods, and where $n$ is the depth of the circuit.

Now, consider an individual cell, $x_0$.
By examining the dependencies of the individual local operations which make up $F$, we can find a range of cells, $P=\{x:a\leq x\leq b\}$ for some $a,b\in\mathbb{Z}$ such that the new state of the cell $x_0$ depends only on the current state of the cells in $P$.
This value must be the value of the quantum state in the cell to the immediate left of $x_{0}$ before $F$ is applied.
We may also find a minimal set of local operators from $F$ such that the new value of the quantum state at cell $x_{0}$ is computed without violating any of the dependency relationships between the local operators.
Without loss of generality, we can implement $F$ by applying these gates first, to obtain the new state of $x_{0}$, and then applying the rest of the gates.
However, by construction, note that none of the operations remaining can affect cells on both sides of $x_{0}$, since these would have already been performed.
This means that the remaining operations can be divided into two operations acting independently on either side of $x_{0}$.
Since, at this point, only cells in the set $P$ have been affected at all, and the cells to the left and right of $x_{0}$ may no longer interact with each other, the cells to the right of $x_{0}$ must contain all of the information needed to update not only their own states, but the state of the cell $x_{b+1}$, directly to the right of $P$.
This is not possible.

This is not to say that a shift-right CA cannot be intrinsically simulated by a reversible CCA using a larger state space.
However, it reflects the fact that it is not physically possible to implement a closed physical system in which all data is shifted to the right on a one-dimensional lattice.
This means that there exist reversible CA whose physical evolution does implicitly depend on interaction with an external environment.
In other words, reversibility is a necessary condition, but not sufficient, to having a truly closed evolution.

%=====================================================================
\section{Construction Techniques}\label{sec:Con}
%=====================================================================

We have seen how CCA and reversible CCA can be constructed as devices which simulate a corresponding CA or reversible CA.
In this section, we present two alternative techniques for constructing translation commutative local interaction functions.

The first construction is a generalization of the idea of partitioned cellular automata, due to Margolus\cite{cabook}.
We consider a CCA in which each cell is divided into two registers, with $\Sigma=\Sigma_1\times\Sigma_2$.
We will consider the second register, with states $\Sigma_2$, as the \emph{control register}.
The idea is to use the control register to determine which non-trivial operations which are being applied to various neighbourhoods.
We must also ensure that these neighbourhoods do not intersect, so that the resulting local interaction function is translation commutative.

We can formalize this idea by considering configurations of the control register.
Define a \emph{region} of the lattice to be a finite subset $R\subset L$, and define a configuration of a region to be an assignment of states in $\Sigma_2$ to the cells of $R$.
Note that we may also consider arbitary translations $x+R$ of any region, for all $x\in L$.
Finally, we say that two regions $R_1$ and $R_2$ intersect non-trivially if $R_1\neq R_2$ and $R_1\cap R_2$ is non-empty.
\begin{definition}
Suppose we are given two finite regions $R_1$, $R_2$, and respective configurations $C_1$, $C_2$.
We say that $C_1$ and $C_2$ are \emph{intersectable} if for some $x\in L$, the $R_1$ and $x+R_2$ intersect non-trivially and there exists and there exists a configuration $C'$ of $R_1\cup(x+R_2)$ which is consistent with both $C_1$ on $R_1$ and $C_2$ on $x+R_2$.
\end{definition}
With this definition, we define a set of configurations to be \emph{compatible} if no two configurations are intersectable.
Given a set $S$ of compatible configurations, and a configuration of the entire lattice, we can find the set of all regions of the lattice which are consistent with a configuration from $S$.
Furthermore, these distinct regions will not intersect each other.

For each configuration $C_i$ over a region $R_i$ in this set $S$, we may consider a function $f_i$ which applied to the $\Sigma_1$ register of each cell in $R_i$ when the corresponding configuration in the $\Sigma_2$ register matches $C_i$.
Because no non-trivial operations are ever applied to two intersection regions, this local interaction function is translation commutative.
The corresponding local update function may be any single-cell operation over $\Sigma$.
In particular, it may later the state of the $\Sigma_2$ register of a cell.

As an example, we may consider the partitioned cellular automaton introduced by Margolus\cite{cabook}.
In this partitioned CA over a one-dimensional lattice $L$, we are given 
two distinct partitionings of the lattice into pairs of cells.
On even time steps, a two-cell operation $U$ is applied according to one partitioning.
On odd time steps, another two-cell operation $V$ is applied according to the other partitioning.
We may implement this by using by adding a register to each cell with alphabet $\Sigma_2=\{0,1,2,3\}$, and by using $\{01,23\}$ as the set of compatible configurations.
For each consecutive pair of cells, the local interaction function applies $U$ to the $\Sigma_1$ register when the $\Sigma_2$ register is in the configuration $01$, and $V$ respectively for the configuration $23$.
By initializing the $\Sigma_2$ registers of the lattice with alternating $0$ and $1$ states, and using a local update function which maps $x\in\Sigma_2$ to $4-x$, the evolution of the resulting CCA simulates the evolution of the given partitioned CA.
Note that while it is possible to construct initial configurations which do not correspond to the evolution of a proper partitioned CA, the evolution of the CCA itself remains well-defined.

Another useful technique for constructing valid translation commutative local transition functions for CCA is that of cell colouring.
This technique begins with a periodic labelling of the lattice cells with colours drawn from a finite set $K$.
The colouring must satisfy the additional property that the colour of each cell must be different from that of each of its neighbouring cells.
The Coloured CA model can be used to construct automata which simulate physical phenomena such as spin chains \cite{lloyd93,benjamin1,cp07}.

At each time step, a particular colour is chosen, and we give a colour update rule which alters only cells of the chosen colour, using the states of its neighbouring cells.
This sequence of colour update rules must eventually repeat.
By using only reversible colour update rules, we may also make reversible Coloured CA.
As an example, we may consider a two-colour, black and white, one-dimensional CA, with a symmetric neighbourhood of radius one. 
At odd time steps white cells are updated with a rule that depends only on its current state, and the state of its two black neighbours.
At even time steps, black cells get similarly updated.

From the construction, we can see that the sequence of colour update rules of a Coloured CA can be implemented as a translation commutative interaction function by adding a \emph{clock} register to each cell which keeps track of the current colour update rule.
It is also possible to show that any CCA can be simulated by an appropriate Coloured CA.

%=====================================================================
\section{Quantum Cellular Automata}\label{sec:QCA}
%=====================================================================

The ideas leading to the construction of Closed CA or, more specifically, reversible CCA have been used as the basis of a new model of quantum cellular automata\cite{cp07}.
Since quantum mechanics deals with the phsyical effects which are significant at the nanoscale, it is important to consider quantum effects when looking into building nanoscale devices\cite{cp06}.
Furthermore, it opens the possibility of quantum computation using such a quantum cellular automata device, applying only global control.
While an introduction to quantum information is outside of the scope of this paper, we wish to summarize some of the known results about quantum CA from \cite{cp07} which arise from the Closed CA model introduced in this paper.

We should note that other models of quantum cellular automata have been proposed, including some earlier models by Zeilinger\cite{grossing88}, Watrous\cite{watrous}, and Meyers\cite{meyer96} and more recent ones, such as that of Schumacher and Werner\cite{sw04}.
These models are not without merit.
However, they all do share a common drawback alluded to in the introduction. 
They all use the traditional mathematical model of CA---which, as we have shown, is not an accurate description of closed physical systems---as the starting point for quantization.
The common result is, understandably, that these models that do not, in fact, represent actual quantum systems.
In particular, all of these models permit a ``Shift-Right'' QCA, which we have shown is not possible in a closed physical system (as proven above for classical closed CA, and in \cite{cp07} for quantum systems).
There are other examples of non-physical behaviour within the earlier models, which have been pointed out before \cite{sw04,cp07}.

By comparison, the quantization of the CCA presented here can be proven to model any and all appropriate quantum phenomena, \emph{and nothing more}. 
In other words, any formal QCA in this axiomatic system can be implemented as a real physical device \cite{cp07}.
Another advantage of using a model of quantum CA which is based on the CCA model, is that it allows construction techniques analogous to those in Section \ref{sec:Con} to be used.

Briefly, a quantum CA can be described as follows.
First, the $d$-state classical alphabet $\Sigma$ of a CCA is replaced with quantum $d$-level system described by a Hilbert space.
Each cell of the lattice contains a physical system, called a \emph{qudit}, which is an instance of this $d$-dimensional Hilbert space.
This leads to the definition of the \emph{Local Unitary Quantum Cellular Automata} (LUQCA), which is a quantum-mechanical analog of the reversible CCA.
Since closed quantum systems evolve according to unitary operations in discrete time steps,  unitary operations are used in the evolution of the quantum CA, corresponding to reversible operations in the reversible CCA.
We can give an analogous definition of translation commutativity for unitary operators.
\begin{definition}
A Local Unitary Quantum Cellular Automaton is a 5-tuple $(L,\Sigma,\mathcal{N},f_{0},g_{0})$ consisting of a $d$-dimensional cell lattice $L=\mathbb{Z}^{d}$, a finite set $\Sigma$ of $d$ orthogonal basis states, a neighborhood scheme $\mathcal{N}$, a unitary translation commutative local interaction operator $f:\left(\mathcal{H}_{\Sigma}\right)^{\otimes\mathcal{N}}\rightarrow\left(\mathcal{H}_{\Sigma}\right)^{\otimes\mathcal{N}}$, and a unitary local update operator $g:\mathcal{H}_{\Sigma}\rightarrow\mathcal{H}_{\Sigma}$.
\end{definition}
Here, $\mathcal{H}_\Sigma$ denotes the complex inner product space spanned by the orthonormal basis given by the states of $\Sigma$.
Note that while there do exist mathematical techniques to handle the uncountable-dimensional Hilbert space that emerges from this definition\cite{neumanninfinite}, for the purposes of simulation the evolution of any finite subregion using a quantum circuit, these need not be considered.

In $\cite{cp07}$, we show that there exists a universal LUQCA that can efficiently and exactly simulate any quantum computation in the quantum circuit model.
We also show that it is possible to build an efficient quantum circuit that simulates the evolution of any finite subregion of the lattice for any finite period of time.
Finally, and most importantly, we show that for any closed quantum physical system consisting of a lattice of identical quantum subsystems with a (Hamiltonian) evolution that his homogeneous in both time and space, there exists a LUQCA that efficiently simulates this system inexactly, but to within an arbitrary precision.
This final result allows us to consider LUQCA as discretized versions of homogeneous physical systems.

%=====================================================================
\section{Conclusion}
%=====================================================================

The main goal of this paper has been to rephrase cellular automata as a model of nature, rather an abstract model of computation.
We showed that the more traditional definition of CA is not well suited for such a role.
In particular, in Section \ref{sec:Rev} we showed that there exist behaviour that is allowed by the traditional model of CA (even traditional reversible CA) that is impossible in a proper closed physical system.

The model we introduce, closed cellular automata or CCA, does not have this drawback.
Hence, it is suitable for the goal we have set out to accomplish. 

We also discussed important results pertaining to the relationship between CCA and the traditional formalism of CA.
In particular, we show that there is an effective procedure to turn any traditional CA into a CCA. 
Likewise, any reversible CA can be turned into a reversible CCA.
The transformations are efficient, but may incur a cost in resources.
This was to be expected, since with CCA all resources are always accounted for, while in traditional CA they are not.
We also discuss the easier result that CCA are a subset of traditional CA.
These results close the gap, and allow the wealth of results pertaining to CA to apply to our new model.

We also discussed techniques for constructing instances of CCA. 
These are useful in both converting traditional CA algorithms to this new model, and for creating algorithms in the CCA formalism directly.

Finally, we briefly discussed one of the main justifications for this body of work: building a foundation upon which a sensible quantization of CA can be built.
The quantization proper is the subject of a different paper \cite{cp07}, however, and we point the readers there for a complete discussion of it.

In short, we have provided a model that abstracts physical phenomena---natural occurring and otherwise---that appeals to our intuition as behaving like cellular automata; furthermore, we have discussed the relationship of this model to the traditional mathematical formalism.
We use it, also, to prove a result about what \emph{cannot} be done by an actual physical system with CA behavior. 
Finally, we discuss how we can use this formalism as a starting point for a quantum model of CA. 
Although we have discussed CA in this paper, this approach can also be used on other traditional models of computation.
It remains an open question whether doing so can produce results interesting results, as it has done in this case.

The authors would like to thank the following agencies for support during the preparation of this manuscript: DTO-ARO, CFI, CIFAR, and MITACS.  C. A. P\'{e}rez-Delgado would also like to thank QIPIRC.

\bibliography{qca,qca2}

\end{document}